# Expanding Advanced Civilizations in the Universe

**CLAUDIUS GROS**
*Department of Physics, Box 151150, University of the Saarland, 66041 Saarbrücken, Germany.*
Email: gros04@lusi.uni-sb.de

The 1950 lunch-table remark by Enrico Fermi 'Where is everybody' has started intensive scientific and philosophical discussions about what we call nowadays the 'Fermi paradox': If there had been ever a single advanced civilization in the cosmological history of our galaxy, dedicated to expansion, it would have had plenty of time to colonize the entire galaxy via exponential growth. No evidence of present or past alien visits to earth are known to us, leading to the standard conclusion that no advanced expanding civilization has ever existed in the milky-way. This conclusion rest fundamentally on the ad-hoc assumption, that any alien civilizations dedicated to expansion at one time would remain dedicated to expansions forever. Considering our limited knowledge about alien civilizations we need however to relax this basic assumption. Here we show that a substantial and stable population of expanding advanced civilization might consequently exist in our galaxy.

## 1. Introduction

Many visions of human settlement of our solar system, or even of the entire milky-way, foresee a rapid expansion via multiplication of human colonies by exponential growth. These kind of scenarios are motivated in part by the empirical observation, that life-forms on earth tend to expand until every habitable niche is occupied. It was taken for granted, that also Homo Sapiens would have a kind of innate drive to multiply and to expand. The advent of dramatically falling birth-rates in industrialized human societies in the last decades of the 20th century has led many to reconsider this stance and reasons for the settlement of outer space are seen nowadays more in terms of intellectual and emotional challenges, than in the need to alleviate population pressure on earth.

There is a growing consensus that human settlement of the solar system might be technically feasible in principle and sociologically possible. Even so, it is still highly controversial in public discussions whether it will ever be desirable to do so. Sociological considerations become even more important when the time-scales for the settlement of extra-solar planets, including interstellar traveling times and the (possible) need to terraform potential extra-solar planets, are considered. Estimates range from optimistic $10^3$ years to somewhat more reasonable $10^5$ years to establish a single new human colony on an extra-solar planet. Minimal estimates to settle a substantial fraction of our galaxy are typically of order $10^6$-$10^7$ years. In any case, a precondition for intra-galactic expansion is the formation of human societies dedicated to the settlement of outer space over very extended periods of time.

Our limited knowledge of the long-term dynamics of human, post-human and extra-solar advanced civilizations does not allow us presently to give a definitive answer to the question, whether humanity will ever produce social organizational forms stable over $10^3$-$10^6$ years on a technically highly advanced level. And once it does, will this society be dedicated, at least partly, to long-term and peaceful expansion? This question is highly non-trivial in view of the present-day low birth-rates in the technological most advanced human societies on earth. One may also argue, from a system-theory point of view, that coexistence in one and the same society of a stable social fabric and the desire to expand over very prolonged periods of time might be exceedingly difficult to achieve.

In any case, we cannot take something for granted for the society of an advanced alien species what we do not know for sure about human societies. We therefore need to consider the possibility that advanced alien civilizations might be dedicated to expansion, but only for limited periods of time. Here we discuss the consequences of this presumption on Enrico Fermi's consideration on the occurrence of expanding advanced civilizations in our galaxy.

## 2. Dynamics of Advanced Civilizations

We will consider here the development of the population of technologically advanced civilizations in the universe and formulate rate equations for the density of advanced civilizations per galaxy. This density may or may-not be substantial. It still makes sense to consider the density of advanced civilizations per galaxy if this density if of the order of about one per galaxy or somewhat smaller. Which would imply, that we would be the only such civilization in our milky-way. The analysis presented here will fail only in the extreme limit that humanity is alone in the entire universe.

We denote with (E) the density of advanced civilizations in our galaxy dedicated towards expansion and with (S) the density of stagnant civilizations. Due to lack of prior knowledge we need to consider the general case that a civilization might change its character over time. The rate equations

$$\dot{E} = (g - e_E - c_E)E + c_s S + b_E$$
$$\dot{S} = c_E E - (e_S + c_S)S + b_S \quad (1)$$

then govern the population dynamics of civilization in our galaxy. Considering $E$ and $S$ as suitable time-averaged quantities we note that Eqs. (1) also apply if only a small number of advanced civilizations exist simultaneously at any given time.

The parameters $b_{E,S} \geq 0$ denote the effective birth rates and are given by the Drake equation (see Ref. 1). The rare-earth scenario [2], the phase-transition scenario [3, 4] and others [1, 5] entail that humanity is alone in the universe. And that humanity has been alone ever since the big-bang, implying $b_{E,S} \to 0$. Here we assume at least one of them to be finite.

$e_{E,S} > 0$ denote the respective extinction rates. Doomsday theories [1, 6] propose that technologically advanced civilizations are bound to destroy themselves in one way or another (or to be destroyed by something else), implying a large value for the extinction rates. $c_E$ is the rate by which a formerly expanding civilization changes character and enters a period of stagnation and $c_S$ the reverse process. A finite growth-rate $g > e_E$ distinguishes an expanding from a stagnant civilization.

Fermi implicitly assumed $c_{E,S} \equiv 0$ and his 1950 lunchtable remark can be cast into the statement: No single advanced civilization with an effective positive growth rate, $g - e_E > 0$, has ever existed in cosmological history.

The solutions of the rate Eqs. (1) grow/decay exponentially like $\sim \exp[\lambda_{1,2} t]$ with eigenvalues

$$\lambda_{1,2} = \frac{g - e_E - c_E - e_S - c_S}{2}$$
$$\pm \sqrt{\left(\frac{g - e_E - c_E + e_S + c_S}{2}\right)^2 + c_E c_S}$$

We may now generalize Fermi's statement for the case $c_{E,S} > 0$: If there are finite birth-rates $b_{E,S} > 0$ for advanced civilizations then the net expansion-rate

$$\text{Max}[\lambda_{1,2}]$$

must be negative due to the fact that no alien civilization has visited or contacted us. This is possible when $c_E$ is large enough. That is, when advanced civilization remain, in the average, in prolonged phases of stagnation with intermittent periods of expanding activities. There is no argument know presently to humanity which would rule out this sort of long-term dynamics for the population of galactic civilizations.

It is then reasonably to assume, when the net expansion rate $\text{Max}[\lambda_{1,2}]$ is negative, that the population of advanced civilizations in the universe settles in a steady state with $\dot{E} = 0 = \dot{S}$. The equilibrium densities $E_0/S_0$ of expanding/stagnant civilizations in the universe are then given by

$$E_0 = \frac{c_S b_S + (e_S + c_S) b_E}{\lambda_1 \lambda_2}$$

and a corresponding expression for $S_0$. Values for $E_0$ and $S_0$ much smaller than one per galaxy are possible (when $b_{E,S} \to 0$) and would imply that humanity is essentially isolated. The magnitude of the parameters entering (1) are, however, highly controversial and we will not attempt here to estimate them.

The crucial point is, on the other side, that a population-density of advanced civilization substantially bigger than one per galaxy is perfectly possible as well and does not contradic logic. Large densities of advanced civilizations might occur also for small birth rates $b_{Es}$ when the absolute magnitude of one of the eigenvalues, $|\lambda_{1,2}|$, is very small. That is, when the average life-time of an advanced civilizations is very long.

In our analysis we have treated the colonies of an expanding civilization as separate identities. They may, or may-not remain committed to expansion long enough to establish, on the average, new colonies of their own. Both cases are covered by the rate

Eqs. (1), which are generally valid. In the first case at least one of the eigenvalues $\lambda_{1,2}$ would be positive and we would run into the Fermi-paradox. In the second case the population of alien civilizations would settle into a stable equilibrium and could be, in principle, substantial.

A finite density of expanding civilizations does not automatically imply an exponentially growing number of alien civilizations. To derive this result we did not need to postulate that all expanding civilizations cease expansion after a certain period. All we have done here is to drop the (ad-hoc) assumption that the colonies of expanding civilizations automatically inherit all their characteristics. There may perfectly well be some exceptional alien civilizations committed to expansion over cosmological time-scales. Theses special civilizations would not be covered by the rate Eqs. (1), they would spin-off continuously new colonies. They would, however, not contribute to the Fermi-paradox, as the number of colonies from these special civilizations would increase only linearly with time.

## 3. Conclusions

How robust is our conclusion? We have basically no knowledge about extraterrestrial civilizations and all we can do are educated guesses. With this caveat in mind it is worthwhile to point out, that human societies tend to change their character considerably even within relative short time-frames, let's say a few thousand years. At least this has been so historically and the equivalent statement applies to human colonies established in the past on earth. An important and yet unsolved issue, with direct relevance to the Fermi paradox, in sociology and political science is in this context whether human societies stable over very long periods of time and dedicated at the same time continuously to exploration and expansion might exist in principle.

\* \* \*